\documentclass[aps,prx,twocolumn,showpacs,superscriptaddress,longbibliography,floatfix,nofootinbib]{revtex4-2}

\usepackage{amsmath, amsfonts, amssymb, bm}
\usepackage{mathtools}
\usepackage[pdftex]{graphicx}
\usepackage{import}

\usepackage[breaklinks,hypertexnames=false]{hyperref}
\hypersetup{colorlinks,linkcolor={magenta},citecolor={blue},urlcolor={blue}} 

\usepackage{braket}
\usepackage{lipsum}
\usepackage{xcolor}

\usepackage[capitalize]{cleveref}

\usepackage{glossaries}
\glsdisablehyper

\newacronym{mbs}{MBS}{Majorana bound state}
\newacronym{abs}{ABS}{Andreev bound state}
\newacronym{fi}{FI}{ferromagnetic insulator}
\newacronym{jj}{JJ}{Josephson junction}
\newacronym{cpr}{CPR}{current-phase relation}

\usepackage[margin=0.9in]{geometry}
\usepackage[utf8]{inputenc}
\usepackage[inline,shortlabels]{enumitem}
\usepackage{longtable}
\usepackage[caption=false]{subfig}
\usepackage{multirow}

\usepackage{sidecap,tikz}
\definecolor{lime}{HTML}{A6CE39}
\DeclareRobustCommand{\orcidicon}{\hspace{-1mm}
	\begin{tikzpicture}
		\draw[lime, fill=lime] (0,0) 
		circle [radius=0.16] 
		node[white] {{\fontfamily{qag}\selectfont \tiny \,ID}};
		\draw[white, fill=white] (-0.0525,0.095) 
		circle [radius=0.007];
	\end{tikzpicture}
	\hspace{-3mm}
}
\foreach \x in {A, ..., Z}{\expandafter\xdef\csname orcid\x\endcsname{\noexpand\href{https://orcid.org/\csname orcidauthor\x\endcsname}
		{\noexpand\orcidicon}}
}

\newcommand{\up}{\uparrow}
\newcommand{\dw}{\downarrow}
\newcommand{\e}{\mathrm{e}}

\newcommand{\si}{\hat{\sigma}_{0}}

\newcommand{\sy}{\hat{\sigma}_{2}}
\newcommand{\sz}{\hat{\sigma}_{3}}

\newcommand{\ti}{\hat{\tau}_{0}}

\newcommand{\ty}{\hat{\tau}_{2}}
\newcommand{\tz}{\hat{\tau}_{3}}

\newcommand{\mbf}[1]{\mathbf{ #1 }}

\newcommand{\pf}{\textrm{Pf}}
\newcommand{\sgn}{\textrm{sgn}}

\begin{document}

\title{Topological superconductivity in a magnetic-texture coupled Josephson junction}

\author{Ignacio Sardinero\orcidC{}}
\affiliation{Department of Theoretical Condensed Matter Physics, Condensed Matter Physics Center (IFIMAC) and Instituto Nicol\'as Cabrera, Universidad Aut\'onoma de Madrid, 28049 Madrid, Spain}

\author{Rubén Seoane Souto\orcidB{}}
\affiliation{Department of Theoretical Condensed Matter Physics, Condensed Matter Physics Center (IFIMAC) and Instituto Nicol\'as Cabrera, Universidad Aut\'onoma de Madrid, 28049 Madrid, Spain}
\affiliation{Division of Solid State Physics and NanoLund, Lund University, S-221 00 Lund, Sweden}
\affiliation{Center for Quantum Devices, Niels Bohr Institute, University of Copenhagen, DK-2100 Copenhagen, Denmark}
\affiliation{Instituto de Ciencia de Materiales de Madrid (ICMM), Consejo Superior de Investigaciones Científicas (CSIC),
Sor Juana Inés de la Cruz 3, 28049 Madrid, Spain.}

\author{Pablo Burset\orcidA{}}
\affiliation{Department of Theoretical Condensed Matter Physics, Condensed Matter Physics Center (IFIMAC) and Instituto Nicol\'as Cabrera, Universidad Aut\'onoma de Madrid, 28049 Madrid, Spain}

\date{\today}

\begin{abstract}
Topological superconductors are appealing building blocks for robust and reliable quantum information processing.
Most platforms for engineering topological superconductivity rely on a combination of superconductors, materials with intrinsic strong spin-orbit coupling, and external magnetic fields, detrimental for superconductivity. 
We propose a setup where a conventional Josephson junction is linked via a magnetic-textured barrier. Antiferromagnetic and ferromagnetic insulators with periodically arranged domains are compatible with our proposal which does not require intrinsic spin-orbit or external magnetic fields. We find that the topological phase depends on the magnitude and period of the barrier magnetization. The superconducting phase controls the topological transition, which could be detected as a sharp suppression of the supercurrent across the junction.
\end{abstract}

\maketitle


\textit{Introduction.---} 
\glspl{mbs} are charge neutral, zero-energy quasiparticle excitations appearing at the boundaries of topological superconductors~\cite{LeijnseReview, Alicea_2012, Aguado_review2017, LutchynReview, prada2020, Marra_JAP2022}. A pair of \glspl{mbs} localized at the ends of a one-dimensional topological superconductor encodes a nonlocal fermionic state~\cite{Kitaev_PU2001}. These states are robust against local perturbations and display non-Abelian exchange properties~\cite{Aasen_PRX2016,BeenakkerReview_20}, making them attractive for fault-tolerant quantum information processing~\cite{Sarma_NPJ2015}. 
The experimental realization of topological \glspl{mbs} requires the combination of superconductivity, helical electrons usually created from spin-orbit coupling, and time-reversal breaking from magnetism~\cite{Flensberg_NRM2021}. Over the last decade~\cite{Zhang2019}, several material platforms have been explored based on topological insulators~\cite{fu2008,Bocquillon2017}, semiconductor nanowires~\cite{Oreg2010, Lutchyn2010, Mourik_Science2012,prada2020}, planar Josephson junctions~\cite{Pientka2017,Hell_PRL2017,Fornieri_Nature2019,Ren_Nature2019}, chains of magnetic adatoms~\cite{Perge_PRB2013,Pientka_PRB2013,Nadj-Perge2014}, and, recently, ferromagnetic insulators combined with other time-reversal symmetry breaking effects~\cite{Sau_PRL2010,manna2020,maiani2021,escribano2021,Liu_PRB2021,Woods_PRB2021,Khindanov_PRB2021,Langbehn_PRB2021,Poyhonen_SciPost2021,Shabani_2021,vaitiekenas2021,Escribano_NPJ2022,vaitiekenas2022,Razmadze_PRB2023}. 
%

\begin{figure}[b!]
 \centering
    \includegraphics[width=1\columnwidth]{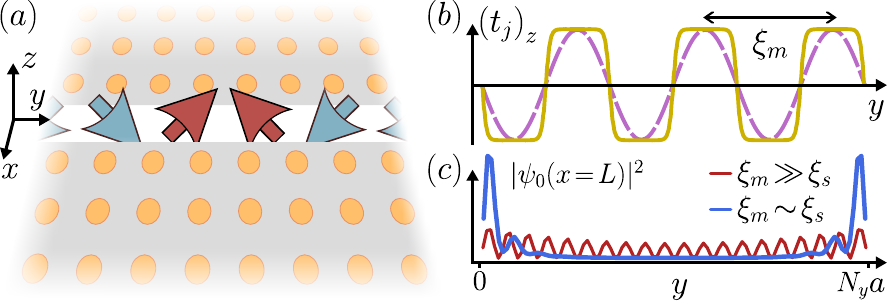}
 \caption{Josephson junction mediated by a magnetic-textured barrier. 
 (a) Two superconductors are linked by a magnetic-texture barrier with a local magnetization that changes in space, schematically represented by arrows. (b) Two possible magnetic texture profiles. (c) Localization of the ground state wavefunction at the interface in the trivial (red) and topological (blue) regimes. 
 }
 \label{fig:setup}
\end{figure}

An alternative strategy has been recently implemented where the spin-orbit is synthetically engineered using spatially-varying magnetic fields~\cite{Kontos_2019,Yazdani2019,Egger_2012_PRB,Oreg_2020,Steffensen_2022_PRR}. 
In proximitized one-dimensional (1D) systems, the spatial magnetic modulation can be achieved by interactions~\cite{Braunecker2010, Braunecker_2013, Klinovaja_2013_PRL}, adatoms~\cite{Beenakker_2011,Franz_2013,Ojanen_2014,Heimes_2014,Xiao2015,Paaske_2016,Paaske_2016b,Cuoco_2017,Kim2018}, or local magnets~\cite{kjaergaard2012, Klinovaja2012, Klinovaja_2013_PRX,Kornich_2020,Frolov_2021}. 
Planar setups offer more sophisticated magnetic textures in proximity to superconductors~\cite{Nakosai_2013,Schnyder_2015,Bena_2015,Fatin_2016,Virtanen_PRB2018,Livanas2019,Matos-Abiague_2019,Zhou_2019,Morr_2020,Xiao_2020,Turcotte_2020,Kotetes_2021,livanas2021,livanas2022,Chatterjee2024}. For example, skyrmion textures on superconductors~\cite{Balatsky_2016,Morr_2021,Mohanta2021,Hoffman_2021} have been recently measured~\cite{Bergmann_2020,Panagopoulos_2021} and
signatures consistent with \glspl{mbs} where found in proximitized magnetic monolayers~\cite{Palacio-Morales2019,Menard2019,Kezilebieke2020}. 
%
Proximitized structures need to be carefully engineered so that the competing magnetic and superconducting orders coexist. 
This challenge could be circumvented if the magnetic texture featuring synthetic spin-orbit interaction is coupled to superconducting contacts in a Josephson setup~\cite{Kontos_2019}. 
Magnetic textures with spatial variation across the junction, i.e., from contact to contact~\cite{Egger_2012_PRB}, have been implemented~\cite{Kontos_2019}. However, for future braiding applications a higher degree of control over the emerging \glspl{mbs} could be achieved with textures along the junction interface, see \cref{fig:setup}. 

Here, we explore such a configuration studying the topological properties of a two-dimensional (2D) \gls{jj} coupled through a magnetic-textured barrier [\cref{fig:setup}(a)] with a spatial modulation along the junction interface [\cref{fig:setup}(b)]. 
We find that the system enters the topological phase when the superconducting coherence length and the magnetization periodicity are comparable. 
The topological regime, characterized by \glspl{mbs} localized at the edges of the \gls{jj} [\cref{fig:setup}(c)], can extend up to rather large occupations and is very sensitive to the phase bias across the junction. 
Moreover, we show that the formation and localization of pairs of \glspl{mbs} has an observable effect on the junction current-phase relation. 
The superconducting phase difference is a substitute of the external magnetic field and helps reaching the topological regime~\cite{Pientka2017,Hell_PRL2017,Lesser_JoPD2022}. Recent experiments have already shown the possibility of tuning subgap states varying the phase difference between planar Josephson junctions~\cite{Fornieri_Nature2019,Ren_Nature2019,Ke_NatComm2019,Dartiailh_PRL2021,Banerjee_2023,Banerjee_2023b}. 
%
Our proposal is thus highly controllable, does not require external magnetic fields, and bypasses the need for intrinsic spin-orbit coupling and low-carrier densities. 

\textit{Model and formalism.---} 
We consider a \gls{jj} formed by two conventional singlet $s$-wave superconductors joined by a magnetic-textured barrier, see \cref{fig:setup}. 
We model the system using a square-lattice tight-binding Hamiltonian $H = H_L + H_R + H_t$, with
\begin{align}
 H_{L,R} ={}& \sum_{\sigma=\up,\dw} 
 (
 -t\sum_{ \langle x, x' \rangle } c_{i'j',\sigma}^\dagger c_{ij,\sigma} - \sum_{i,j} \mu_{ij} c_{ij,\sigma}^\dagger c_{ij,\sigma} 
 )
 \nonumber \\  \label{eq:hlr}
 & + \sum_{i, j} \Delta_0 \e^{i \phi_{L,R}} c^\dagger_{ij,\up} c^\dagger_{ij,\dw} + \mathrm{h.c.}\,, 
\end{align}
being the Hamiltonian of the superconducting leads, where $\langle x,x' \rangle$ stands for nearest-neighbors combinations of the horizontal and vertical indices $i,i'$ and $j,j'$. 
The operator $\hat{c}^\dagger_{ij,\sigma}$ ($\hat{c}_{ij,\sigma}$) creates (annihilates) an electron with spin $\sigma$ on the lattice site $(i,j)$. 
Here, $t$ is the nearest-neighbors hopping integral, $\mu_{ij}\equiv\mu$ the uniform chemical potential of the lattice, $\Delta_0>0$ the superconducting pairing amplitude, and $\phi_{L,R}$ the superconducting phases; we denote their phase difference as $\phi=\phi_R-\phi_L$. 
We consider a finite size lattice with $N_x$ and $N_y$ horizontal and vertical sites, respectively, and only examine symmetric junctions where both superconductors have the same gap, length $L=N_x a/2$ (with $a$ being the lattice distance constant), and width $N_y a$. 

The superconductors couple via a magnetic-textured barrier mediating tunneling between them, 
\begin{equation} \label{eq:vlr}
    H_t = - \frac{1}{2} \sum_{ \sigma, \sigma'}\sum_{j}  t_{j,\sigma\sigma'} \hat{c}^\dagger_{L,j,\sigma} \hat{c}_{L+1,j,\sigma'} + \mathrm{h.c.}\,. 
\end{equation}
We consider that the magnetization of the barrier has a spatial modulation, given by the matrix in spin space
\begin{equation}\label{eq:hopping}
    \hat{t}_{j} = t_0 \si + \bm{t}_j \cdot \bm{\sigma} .
\end{equation}
The index $j$ runs along the width of the junction (\cref{fig:setup}) and $\bm{\sigma}$ is the vector of Pauli matrices $\hat{\sigma}_{0, 1, 2, 3}$ in spin space. Here, $t_0$ is the uniform amplitude for the spin-conserving hopping term and $\bm{t}_j$ the spatially-varying spin-tunneling part. 


\textit{Analytic 1D topological model.---} 
To study the bulk topological properties of the system, we consider a perfect harmonic spatial variation along the spin $yz$-plane with period $ \xi_m $ and constant magnitude $t_\text{m}$, 
\begin{equation}\label{eq:ty-sin}
    \bm{t}_j = t_m \left[0,\sin{(2\pi j a/\xi_m)}, \cos{(2\pi j a/\xi_m)}\right].
\end{equation}
We reach a solvable model setting $N_x=2$, effectively reducing the system to two superconducting linear chains coupled along the $y$-direction by \cref{eq:ty-sin}, and assuming $N_y\rightarrow\infty$, i.e., applying periodic boundary conditions to go to the bulk limit. 
Then, we change into a rotating-frame basis so that the magnetization orientation always falls along the $z$-axis~\cite{kjaergaard2012}. 
As a result, the two linear superconductors acquire an effective spin-orbit coupling~ \cite{Braunecker2010,Klinovaja2012,kjaergaard2012}, and are described by the (left, right) Hamiltonians
\begin{equation}\label{eq:lead_hamil}
\begin{split}
    \hat{h}_{L,R} ={}& \left( -2t \cos(k_y a) - \mu - \dfrac{\pi^2}{\xi_m^2} \right) \si\tz \\ & +  \sin(k_y a)\dfrac{2 \pi i}{\xi_m}  \sy\tz + \Delta_0\e^{i\phi_{L,R}} \sy \ty,
\end{split}
\end{equation}
coupled by the tunnel Hamiltonian
\begin{equation}\label{eq:min-hop}
    \hat{h}_{t} = ( t_0 \si + t_{m} \sz ) e^{i\phi/2} \tz ,
\end{equation}
with $\hat{\tau}_{1,2,3}$ acting in Nambu (particle-hole) space. 

The Hamiltonian of each linear chain $\hat{h}_{L,R}$ is particle-hole and mirror symmetric, while $\hat{h}_{t}$ breaks time-reversal symmetry. Therefore, the system is in BDI class~\cite{Pientka2017,Mizushima_NJP2013} and can be characterized by a  $\mathcal{W} \in \mathbb{Z}$ invariant~\cite{Fulga_PRB2010}, whose parity $\mathcal{M}=(-1)^\mathcal{W} \in \mathbb{Z}_2$ can be determined after a change to the so-called Majorana basis. 
Rewriting $H{(k_y)}=\hat{h}_L+\hat{h}_R+\hat{h}_t$ as a skew-symmetric matrix $A$ is possible by a unitary transformation, and we thus reach 
\begin{equation}{\label{eq:maj}}
 \mathcal{M} = \sgn \left( \pf[A(0)]/\pf[A(\pi/a)] \right), 
\end{equation}
with $\pf[A]$ referring to the Pfaffian of $A$. See Supplemental Material (SM)~\cite{SM} for more details. 
%


\begin{figure}[ht!]
 \centering
 \def\svgwidth{\columnwidth}
\subfloat{\includegraphics[width=1.0\columnwidth]{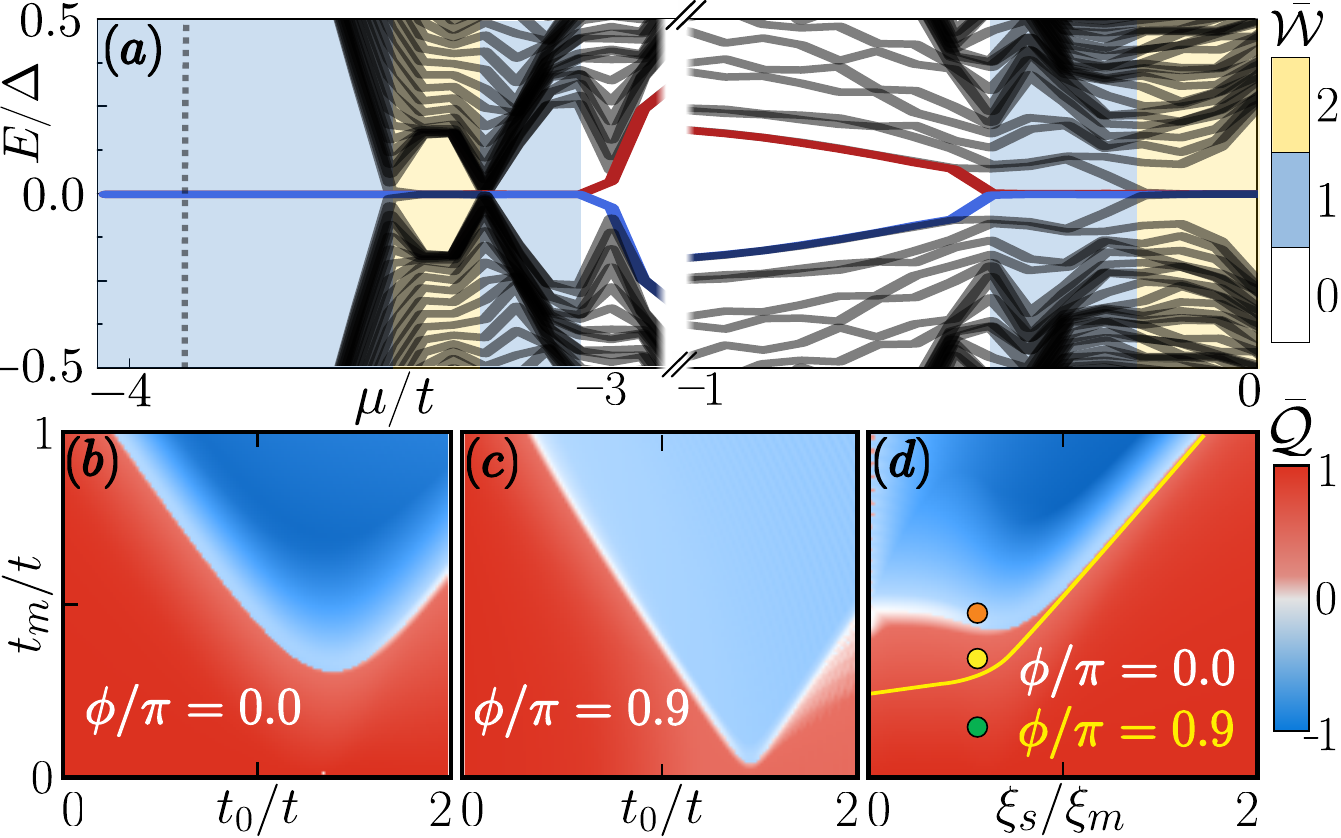}} 
 \caption{Topological phase diagram. (a) Lowest 40 energy bands as a function of the chemical potential for $\xi_m \approx 0.8 \xi_s $ and $ t_m/t = 0.75 $. The coloured background represents the number of pairs of edge states $\bar{\mathcal{W}}$. Figures (b-d) show maps of $\bar{\mathcal{Q}}$ at $\mu/t=-3.8$; blue indicates the topologically nontrivial phase. 
 (b,c) $\bar{\mathcal{Q}}$ with $\xi_m = \xi_s $ as a function of $t_0$ and $t_{m}$ for (b) $\phi=0$ and (c) $0.9\pi$. 
 (d) $\bar{\mathcal{Q}}$ as a function of $t_{m}$ and $\xi_m$ for $t_0/t = 1$ and $\phi=0$. The topological phase boundary for $0.9\pi$ is highlighted as a yellow line. 
 In all cases, $\Delta_0/t = 0.2$ and $N_x \times N_y = 16 \times 128$. }
 \label{fig2}
\end{figure}

\textit{Topological superconductivity.---} 
Using the analytic model as a guide, we now focus on the finite size system depicted in \cref{fig:setup} and described by \cref{eq:hlr,eq:vlr}. 
We first consider a harmonic variation of the magnetic texture with constant amplitude $t_m$, \cref{eq:ty-sin}, although our findings remain qualitatively invariant for other periodic magnetization profiles~\cite{SM}. 
The barrier locally polarizes the tunneling electrons inducing an effective exchange field in the two superconductors. Additionally, we account for a local exchange field close to the barrier~\cite{SM}. 


In the absence of magnetism ($t_m=0$), the system features a set of spin-degenerate Andreev bound states inside the superconducting gap with an energy that depends on the phase and the transmission of the channel~\cite{Beenakker_PRL1991}. 
A uniform spin polarization along a fixed direction, $\mbf{t}_j=t_m(0,0,1)$, splits these subgap states in two different spin species, eventually closing the superconducting gap. 
The system, however, is still in the trivial phase. 
We describe the magnetic texture by introducing the spatial variation in \cref{eq:ty-sin}, which mixes the two spin components facilitating the formation of equal-spin triplet pairing close to the junction~\cite{BergeretRMP2005}, to the second term of \cref{eq:hopping}. 
We now describe the conditions for the gap reopening indicating a transition into the topological phase with localized \glspl{mbs} at the edges of the \gls{jj}. 

To characterize the topological phase of the finite system we define an approximate topological invariant $ \bar{\mathcal{W}} $, which we compute as the number of \glspl{mbs} pairs lying at $E \approx 0$ separated from the subsequent bulk modes by an effective gap $\delta\leq \Delta_0$~\cite{SM}.
The system features two Majorana states ($\bar{\mathcal{W}}=1$) when the chemical potential of the 2D superconductors is $\mu\sim-3.5t$. 
This regime is shown in \cref{fig2}(a) highlighted by a blue background color. 
Other topological phases with $\bar{\mathcal{W}}>1$ appear at higher fillings of the superconductors, shown as a yellow background in \cref{fig2}(a). 
We thus focus below on fillings with only one \gls{mbs} pair. 

We compute $\bar{\mathcal{Q}} =(-1)^{\bar{\mathcal{W}}} \delta/\Delta_0$ in \cref{fig2}(b-d) to characterize the topological phase and show that the magnetic tunneling $t_{m}$ required to enter the topological regime (blue regions) is minimal close to $t\approx t_0$, where the junction's normal transmission is maximal. 
%
%
In the analytic 1D model, the boundary between topological phases for $\phi=0$ [white regions in \cref{fig2}(b-d)] follows the simple condition~\cite{SM}
\begin{equation}\label{eq:phase-bound-1}
    t_m^2 = \left[ \mu_{\rm eff} (\xi_{m}) \pm t_0 \right]^2 + \Delta_0^2,
\end{equation}
with $\mu_{\rm eff} (\xi_{m})= 2t - \mu + \pi^2/\xi_{ m }^2$~\cite{SM}. 
The minimum parameters to reach the topological phase are thus $t_m=\Delta_0$ and $|\mu_{\rm eff}|=t_0$, which qualitatively coincides with the minimum of the blue region of \cref{fig2}(b). 

The phase difference $\phi$ between superconductors provides another way of controlling the topological phase transition. Tuning $\phi$ from $0$ to $\pi$ reduces the energy of the subgap states so that the phase transition occurs for lower $t_m$ values, with a minimum located at $\phi=\pi$ (mod $2\pi$). \Cref{fig2}(c) shows the phase diagram for $\phi=0.9\pi$ illustrating the reduction of the minimal $t_m$ required to enter the topological phase with respect to the $\phi=0$ case shown in \cref{fig2}(b). However, we note that now the topological gaps are smaller (lighter blue in the topological region) than the ones found for $\phi=0$. 
From the 1D model, the minimal required $t_m$ value is given by
\begin{equation}\label{eq:phase-bound-2}
    t^{\rm min}_{m} = \Delta_0 \sqrt{ 1 + \cos{\phi} + \Delta_0^2 \dfrac{ \sin^2{\phi}}{2\mu_{\rm{\rm eff}} (\xi_{m}) } } ,
\end{equation}
with the limits $t_{m}(\phi=0)=\Delta_0$ and $t_{m}(\phi=\pi)=0$~\cite{SM}. Comparing the minima in \cref{fig2}(b) and \cref{fig2}(c) shows that these results are in qualitative agreement with the finite-size calculations. 

The spatial periodicity of the magnetic texture, $\xi_{m}$, is another important parameter for reaching the topological phases, see \cref{fig2}(d). The local magnetization induced in the superconductors decays with a typical length-scale given by the superconducting coherence length $\xi_{s}$~\cite{tokuyasu1988}. 
Therefore, 
for long magnetic periods, $\xi_{s}/\xi_{m}\ll1$, electrons only feel the exchange field locally induced by the barrier and the gap collapses for sufficiently strong $t_{m}$ values. 
In the opposite regime of small magnetic periods, $\xi_s/\xi_m\gg1$, the barrier magnetization cannot close the superconducting gap and the system remains in the trivial phase. 
Consequently, a robust topological phase requires $\xi_{s}/\xi_{m}\sim1$ where the topological gap is maximum while the $t_{m}$ required for the phase transition is minimal [\cref{fig2}(d)].

\begin{figure}
  \includegraphics[width=1.0\columnwidth]{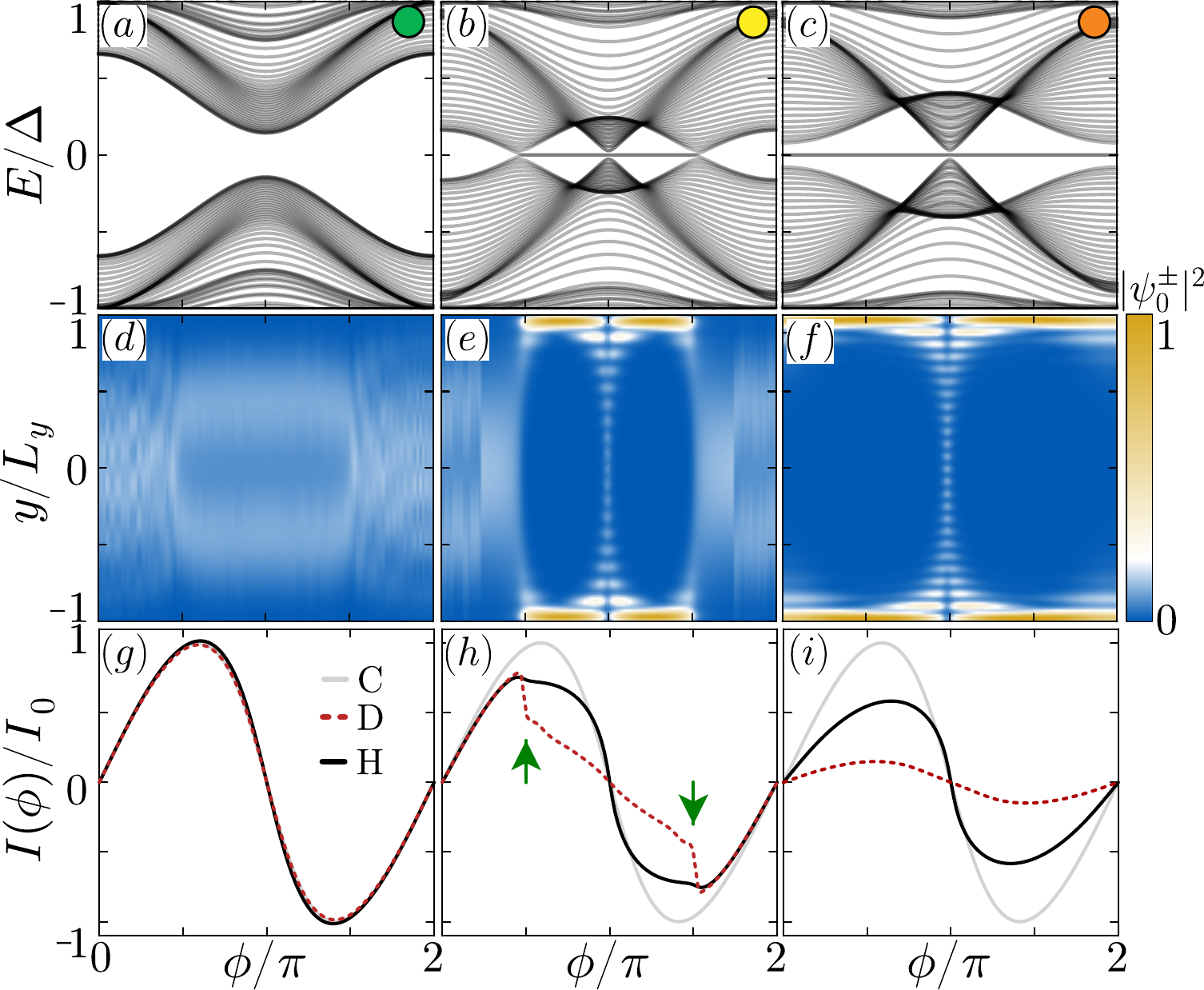}
  \caption{Phase-controllable topological order. \mbox{(a-c)} Phase dependence of energy bands for $t_{m}/\Delta_0=1.0$, $2.0$, and $2.5$ in the barrier with harmonic texture. 
  (d-f) Spatial distribution of the lowest-energy wavefunction $|\psi_0^{\pm}|^2$ evaluated along the junction interface ($i=L$).
  (g-i) Current-phase relation in the case of the clean (C), harmonic- (H) and domain-textured (D) junctions. $I_0$ is the current maximum for the clean junction with $t_{m} = 0$ and $t_0 = t$. Arrows indicate the kink at the phase transition. 
  In all cases, $\mu/t=-3.8$, $\xi_m = 2\xi_s $, and $\Delta_0/t=0.2$. }
  \label{fig3}
\end{figure}

\textit{Phase-controlled topological order.---} 
We now discuss in more detail the effect of the phase difference between superconductors. As shown in \cref{fig2}, the system can transition from the trivial to the topological regime when increasing the phase difference. Therefore, a finite superconducting phase difference reduces the energy of the subgap states, making it possible to transition to the topological regime for smaller $t_m$ values. The superconducting phase is thus a convenient parameter to control topology and, consequently, the emergence and localization of \glspl{mbs}. As an illustration, we compare in \cref{fig3} three different configurations: far into the trivial region (left column); close to the boundary between topological phases from the trivial region (center); and inside the nontrivial phase (right). 
In the first case (left), the modulation of the subgap states with the phase is insufficient to close the gap. The center column shows a gap closure and reopening for a finite phase ($\phi\sim\pi/2$), so that a pair of topological edge states appear when it reopens ($\phi > \pi/2$). 
Figures~\ref{fig3}(d-f) display the ground state wavefunctions, $|\psi_0^\pm|^2$, showing the formation of localized edge modes in the topological regime.
In the nontrivial case (right column), the superconducting phase increases the topological gap in the region away from $\phi=0$ and $\phi=\pi$. At $\phi=\pi$ the gap collapses and Majoranas delocalize across the full system. 

The results presented thus far correspond to a harmonically rotating magnetic texture along the junction interface [dashed line in \cref{fig:setup}(b)]. 
A more realistic texture is represented by the solid line in \cref{fig:setup}(b) featuring anti-parallel domains with non-coplanar domain walls described by
\begin{equation}\label{eq:ty-domain}
    \bm{t}_j = t_m \lambda_0^{-1} \left[0, \tanh\left(\lambda\sin{y_j}\right), \tanh\left(\lambda\cos{y_j} \right)\right] ,
\end{equation}
with $y_j=2\pi j a/\xi_m$, $\lambda_0 = \tanh(\lambda)$ and the parameter $\lambda$ controlling the length of the magnetic domains~\cite{SM}. 
Using this domain magnetization model with the same parameters of \cref{fig2} we reproduce the positions of the crossings in \cref{fig3} and the overall structure of the map is preserved, see Ref.~\cite{SM}. 
Consequently, we conclude that the details of the magnetic texture do not qualitatively affect the topological phase, as long as $\xi_{s}/\xi_{m}\sim1$. 

Finally, we compute the Josephson current at zero temperature, i.e., $I\propto\sum_{\epsilon_i<0}\partial_\phi \epsilon_i(\phi)$, for the negative eigenvalues $\epsilon_i$ of the full Hamiltonian~\cite{SM}. We show the \gls{cpr} for the three cases commented above in \cref{fig3}(g-i). Interestingly, the phase-induced topological transition as a function of the phase appears as a kink on the \gls{cpr}, see arrows in \cref{fig3}(h). 
This kink is produced by an avoided crossing of the remaining trivial Andreev bound states resulting from the topological protection of the newly formed gap, and it is more pronounced for the domain-texture (red dashed lines) than for the harmonic case. 


\textit{Role of the magnetization profile.---}
The domain-texture model, \cref{eq:ty-domain}, allows us to systematically introduce disorder by randomizing the domain size ($\xi_{m}$), domain magnetization ($t_{m}$), or domain wall helicity. 
Weak disorder in $\xi_{m}$ and $t_{m}$ does not disrupt the topological phase because these quantities preserve the system symmetry~\cite{SM}. Indeed, disorder on $\xi_{m}$ makes the topological gap size variable along the interface, since it is controlled by the magnitude of the effective spin-orbit coupling; while random domain magnetization values broaden the value of $\phi$ needed for the phase transition, thus making the avoided crossing in \cref{fig3} wider and the kink less pronounced. 
Conversely, changes in domain wall helicity alter the sign of the effective spin-orbit coupling along the wire, leading to topological phase boundaries where different highly overlapping subgap modes can enter the topological gap~\cite{SM}. 


\textit{Conclusions.---} In this work, we have shown the onset of topological superconductivity in a Josephson junction mediated by a spin-textured barrier. Using a two-dimensional tight-binding model, we identified the conditions for the emergence of topological Majorana zero modes at the edges of the system. 
The presence of the magnetic texture eliminates the need for spin-orbit coupling or external magnetic fields, while the phase bias across the junction provides control over the topological phase. 
We support these results computing the topological invariant in an analytical 1D model of the junction. Additionally, we investigate the impact of disorder on the topological phase transition, finding that only sign changes of the magnetic texture's helicity introduce trivial states within the topological gap. 

Our work proposes a platform for topological superconductors that do not rely on intrinsic spin-orbit coupling and external magnetic fields.
Materials including antiferromagnetic insulators with a small out-of-plane magnetization or ferromagnetic insulators with ordered domains are suitable candidates for magnetic-textured barriers~\cite{Bell_2003_PRB,*Bell_2003_PRB, Kamra_2018_PRL,Lado_PRL2018, Luntama_2021_PRR, Fyhn_2023_PRL,Akash2023,Johnsen2023}. The possible detrimental effect from stray fields could be reduced by choosing a texture with fields pointing away from the superconductors.

\acknowledgments
We thank E.J.H. Lee and A. Levy Yeyati for insightful discussions. 
I.S. and P.B.~acknowledge support from the Spanish CM ``Talento Program'' project No.~2019-T1/IND-14088 and the Agencia Estatal de Investigaci\'on projects No.~PID2020-117992GA-I00 and No.~CNS2022-135950. 
R.S.S.~acknowledges funding from the Spanish CM ``Talento Program'' project No.~2022-T1/IND-24070, the European Union’s Horizon 2020 research and innovation program under the Marie Sklodowska-Curie Grant Agreement No. 10103324, the European Research Council (ERC) under the European Union's Horizon 2020 research and innovation programme under Grant Agreement No. 856526 and Nanolund.


%




\clearpage

\onecolumngrid
\setcounter{equation}{0}
\renewcommand{\theequation}{S\,\arabic{equation}}
\setcounter{figure}{0}
\renewcommand{\thefigure}{S\,\arabic{figure}}
\begin{center}
\large{\bf Supplemental material to ``Topological superconductivity in a magnetic-texture coupled Josephson junction''}
\end{center}



%
%
%
%

This Supplementary Material is organized as follows. 
We introduce the one-dimensional model and study its topology in \cref{sec:app-analytics}. 
In \cref{sec:app-domain-model} we present the domain magnetic texture model and its generalization to include disorder effects. 
Next, we define in \cref{sec:app-aprox-top-inv} the approximate topological invariant valid for the finite-size systems explored in the main text. 
\Cref{sec:app-disorder} is devoted to study disorder effects on all relevant parameters of the domain magnetic texture. 
We then analyze the impact of the band filling on the topological phase in \cref{sec:app-band-fill}. Finally, we define the Josephson current in \cref{sec:app-cpr}.

\section{Analytic 1D topological model}
\label{sec:app-analytics}
We develop a 1D model that can be solved exactly to obtain analytic insights on our main results. The model consists of two infinite 1D superconductors along the $y$ direction, L and R, coupled via a harmonic spin-textured barrier $V_{LR} (y)$. It is represented by the Hamiltonian
\begin{equation}\label{eq:Hamil-1d}
 \check{H} =
 \begin{pmatrix}
  H_L & V_{LR} (y) \\
  V_{LR}^\dagger (y) & H_R
 \end{pmatrix} ,
\end{equation}
which spans in Nambu, spin, and L-R spaces. 
Each superconductor is described as
\begin{equation}
 H_{L,R} = - \left[ \mu + 2t \cos(k_y a) \right] \si \tz + \Delta_{L,R} \e^{i\phi_{L,R}} \sy \ty
 ,
\end{equation}
where the Pauli matrices $\hat{\sigma}_{0,1,2,3}$ and $\hat{\tau}_{0,1,2,3}$ respectively act in spin and Nambu spaces, $\mu$ is the uniform on-site potential, $t$ the nearest-neighbours hopping, and $\Delta_{L,R}\e^{i\phi_{L,R}}$ the superconducting pair potential. We impose that the superconductors are equal, $\Delta_L=\Delta_R\equiv\Delta_0>0$, and denote $\phi=\phi_R-\phi_L$ their phase difference. We consider the system to be translation invariant and, therefore, treat the problem in momentum space with conserved wavenumber $k_y$. The tunneling between the two superconductors is given by
\begin{equation}
V_{LR} (y) =
 \begin{pmatrix}
    -t_0 - t_m \cos\theta(y) & -t_m \sin\theta(y) \\
    t_m \sin\theta(y) & -t_0 + t_m \cos\theta(y)
 \end{pmatrix} \tz,\quad
 \theta(y) = \dfrac{2\pi}{\xi_m} y ,
\end{equation}
where $t_m$ and $\xi_m$ are the amplitude and the spatial period of the magnetic texture, respectively, and $t_0$ a spin-independent hopping. 

To simplify the Hamiltonian, we locally rotate the electron spin basis so that it always points in the $z$ direction using the unitary transformation
\begin{equation}
U = \left[ \cos(\theta/2) \si - i \sin(\theta/2) \sy \right] \ti.
\end{equation}
The rotated Hamiltonian $\tilde{H} = U^\dagger H U$ 
acquires two extra terms coming from the transformation
\begin{align}
 U^\dagger \partial_y^2 U - \partial_y^2 =  U^\dagger U'' \partial_y^2 + 2 U^\dagger U' \partial_y ,
\end{align}
with $U'=\partial_yU$. Substituting $\partial_y^2 \rightarrow -2 t \cos(k_y a)$ and $\partial_y \rightarrow -2 t \sin(k_y a)$ we find
\begin{equation}
 \tilde{H}_{L,R} = H_{L,R} - \dfrac{\pi^2}{2 \xi_m^2} \si \tz - \dfrac{2 \pi}{\xi_m} \sin(k_y a) i\sy \tz 
,
\end{equation}
giving rise to an effective spin-orbit $\alpha_{\rm eff} \equiv \pi/(2 \xi_m)$ and an effective chemical potential $\mu_{\rm eff} \equiv \mu + \pi^2/(2 \xi_m^2)$. Now, the Zeeman term is aligned along the spin $z$ axis and the tunneling term reads as
\begin{equation}
\tilde{V}_{LR} (y) =  U^\dagger {V}_{LR} U =
 - \left( t_0\si + t_m\sz \right) \tz.
\end{equation}

\subsection{Topological classification and topological invariant}

We can now classify the Hamiltonian in \cref{eq:Hamil-1d} according to its symmetries. 
The superconducting terms impose particle-hole symmetry (PHS) but the magnetic texture breaks time-reversal symmetry (TRS). Therefore, one could in principle assume that the system has no chiral symmetry and would belong in symmetry class D~\cite{Schnyder_PRB2008}. 
However, aside from the three non-spatial symmetries (TRS, PHS, chiral), we can also consider a mirror symmetry with respect to the interface. A combination of time-reversal and mirror symmetries, $\tilde{\mathcal{T}} = \mathcal{T} \mathcal{M}$, together with PHS, realizes the chiral symmetry~\cite{Mizushima_NJP2013}. As a result, \cref{eq:Hamil-1d} belongs in the BDI class, explaining the appearance of phases with more than one \gls{mbs} per side (yellow regions of \cref{fig2} in the main text). 

Since the system is two-dimensional and belongs in class BDI, it is possible to take it to an off-diagonal form in the chiral basis and use the winding number $\mathcal{W}$ as topological invariant~\cite{Tewari_2012_PRL}. To do this, we consider the three symmetries that the system has, and how each term in the Hamiltonian transforms under each of them, and write down a possible matrix representation for each symmetry. The global chiral symmetry can then be written as 
\begin{equation}
    \check{C} = - \hat{\tau}_2 \hat{u}_1 \hat{\sigma}_3,
\end{equation}
with Pauli matrices $\hat{u}_{0,1,2,3}$ acting in the left-right space. 
As a result, the Hamiltonian in this basis is
\begin{equation}
 \check{C} \check{H}(k_y) \check{C}^\dagger =
 \begin{pmatrix}
  0 & A(k_y) \\
  A^T(-k_y) & 0
 \end{pmatrix},
\end{equation}
with 
\begin{equation}
    A(k) = 
    \begin{pmatrix}
        \mu_{\rm eff} + 2t\cos{k_y} & -\alpha_{\rm eff} \sin{k_y} & t_0 - t_{m} & -i\Delta_0 e^{i \phi/2}  \\
        \alpha_{\rm eff} \sin{k_y} & \mu_{\rm eff} + 2t\cos{k_y} &  -i\Delta_0 e^{i \phi/2}  & t_0 + t_{m} \\
         t_0 - t_{m} & -i\Delta_0 e^{-i \phi/2}  & \mu_{\rm eff} + 2t\cos{k_y} &  -\alpha_{\rm eff} \sin{k_y} \\
         -i\Delta_0 e^{-i \phi/2}  & t_0 + t_{m} & \alpha_{\rm eff} \sin{k_y} & \mu_{\rm eff} + 2t\cos{k_y}
    \end{pmatrix}.
\end{equation}

The parity of the winding gives the onset of even or odd number of uncoupled MBSs at each edge of the system. Therefore, the invariant we are interested in is given by~\cite{Fulga_PRB2010, Altland_PRB1997}
\begin{equation}\label{eq:topo-inv}
 \mathcal{M} = \sgn \left\{ \dfrac{\pf\left[ \check{H}_{sk}(k_y = 0)\right]}{\pf\left[\check{H}_{sk}(k_y = \pi/a)\right]} \right\} = (-1)^{\mathcal{W}},
\end{equation}
where $\pf\left[\check{H}_{sk}(k_y)\right]$ stands for the Pfaffian of $\check{H}_{sk}$, which is the Hamiltonian rotated into a skew-symmetric form $\check{H}_{sk}^T=-\check{H}_{sk}$; namely $ \check{H}_{sk}(k_y) = U \check{H}(k_y)  U^\dagger$, with
\begin{equation}
    U = \frac{1}{\sqrt{2}}\begin{pmatrix}
        1 & 1 \\ -i & i
    \end{pmatrix}
    \si \hat{u}_0.
\end{equation}
The topological invariant can thus be recast as
\begin{equation}
 \mathcal{M} = \sgn ( c_0 / c_\pi ) = \sgn ( c_0  c_\pi ),
 \label{eq:M}
\end{equation}
with
\begin{equation}
c_{0,\pi} = \Delta_0^4 + 2 \Delta_0^2 (\mu_{\rm eff}^{0,\pi})^2 + (t_0 - t_m)(t_0 + t_m)\cos\phi + \prod\limits_{\substack{s_0 = \pm 1 \\ s_m = \pm 1}} (\mu_{\rm eff}^{0,\pi} + s_0 t_0 + s_m t_m), \label{eq:topo_cond_phi}
\end{equation}
and
\begin{equation*}
 \mu_{\rm eff}^{0,\pi} = \mu + \pi^2/\xi_m^2 \pm 2t.
\end{equation*}

\subsection{Boundaries of the topological phase}
\Cref{eq:M} has sign changes representing the system's topological phase transitions every time the numerator or the denominator go through a zero. Setting $\mathcal{M}=0$ we get eight solutions $\pm t_{m}^{0,\pi;\pm}$ with
\begin{equation}
    \left(t_m^{0,\pi;\pm}\right)^2 = t_0^2 + \mu_{\rm eff}^{0,\pi} + \Delta_0^2 \cos\phi \pm \sqrt{2\mu_{\rm eff}^{0,\pi}\left[2t_0^2-\Delta_0^2\left(1-\cos\phi\right)\right]-\Delta_0^4\sin^2\phi} . 
    \label{eq:fullCond}
\end{equation}
\Cref{eq:fullCond} reduces to a very simple condition for $\phi=0$ indicating the transition into a topological phase:
\begin{equation}
    t_{m}^2 > \left[\pm 2t - \mu_{\rm eff} \pm t_0\right]^2 + \Delta_0^2.
    \label{eq:topo_cond}
\end{equation}

\Cref{eq:topo_cond} describes two circles in the $t-\Delta_0$ plane each one with radius $t_m$ and centered around $ 2|t| = \pm\mu_{\rm eff} - t_0$. 
The system is in the topological phase for parameters inside the two circles where $t_m$ dominates and Eq.~\eqref{eq:topo_cond} is satisfied. Outside, the system is in the trivial regime. The two circles intersect when $t_m>\mu_{\rm eff}$, where there are topological phases with higher winding number hosting more than a Majorana pair. 
The superconducting phase difference $\phi $ enters \cref{eq:topo_cond_phi} in the term $ (t_m^2 - t_0^2) \cos(\phi)$. Therefore, a phase $\phi=\pi$ exchanges the role of $t_{m}$ and $t_0 $ in \cref{eq:topo_cond}, while for $\phi=\pi/2$ they play the same role. 

\subsection{Role of the superconducting phase difference}

From \cref{eq:fullCond} we can deduce the minimal $t_m>0$ solution marking the transition to the topologically nontrivial $\mathcal{W}=1$ phase in which we are interested. The critical values of $t_0$ that minimize that solution are 
\begin{equation}
    t_0^{\rm{min}} = \sqrt{ \mu_\text{eff}^2 + \Delta_0^2\sin^2\frac{\phi}{2} + \dfrac{\Delta_0^4}{4\mu_\text{eff}^2}\sin^2\phi } \, .
    \label{eq:t0_opt}
\end{equation}
\Cref{eq:t0_opt} indicates the horizontal position of the minimum of the parabolic nontrivial region (blue area) in the left panels of \cref{fig:analytic_numeric_comp}, which is in good correspondence with the full 2D calculation, see middle panels. 

Substituting \cref{eq:t0_opt} into \cref{eq:fullCond} we obtain the corresponding minimal magnetic tunneling:
\begin{equation}
    t_{m}^\text{min} = \Delta_0 \sqrt{ \cos^2\frac{\phi}{2} + \left(\dfrac{\Delta_0}{2|\mu_\text{eff}|}\right)^2 \sin^2\frac{\phi}{2} } \, .
    \label{eq:tm_min}
\end{equation}

The phase difference can thus reduce the minimum $t_m$ required to enter the topological phase for a given value of $t_0$. This effect explains the expansion of the topological phase between \cref{fig2}(b) for $\phi=0$ and \cref{fig2}(c) with $\phi\lesssim\pi$ in the main text. 
For $t_m/t\gtrsim1$ the superconducting phase does not significantly change the topological boundary, see, e.g., in \cref{fig2}(d) of the main text the comparison between the numerical finite-size calculation (color map) and the analytical result represented by a yellow line. 

Consequently, the analytical model qualitatively reproduces the behavior at low band fillings of the finite-size numerical calculation described in \cref{fig2} of the main text, see also \cref{fig:analytic_numeric_comp} below. The finite-size effects only introduce small variations in the minimal $t_0$ and $t_m$ values where the topological transition occurs. 

\section{Domain magnetic texture model\label{sec:app-domain-model}}

The domain model depicted in \cref{fig:setup}(b) of the main text consists of an alternating series of regions with parallel magnetization (domains), like in an antiferromagnetic texture, separated by smaller regions where the magnetization rotates from one orientation to its opposite (domain walls). 
We again assume a magnetization with constant magnitude $t_m$ that rotates in the $yz$-plane with period $\xi_m$, like in \cref{eq:ty-sin} of the main text. 
Each period is formed by two domains involving $d$ sites each and two domain walls with $w$ sites each, that is, $\xi_m=2a(d+w)$. 
The two domains in a period feature anti-parallel magnetization along the $z$ spin axis, $\pm t_m \hat{z}$, and the magnetization in the domain walls rotates from $+t_m\hat{z}$ to $-t_m\hat{z}$. 
The domain-model magnetic texture can then be compactly described by [cf. \cref{eq:ty-domain} of the main text]
\begin{equation}\label{eq:ty-domain-app}
    \mbf{t}_j = \frac{t_m}{\tanh\lambda} \left( 0, 
    \tanh\left[\lambda \sin\left(\frac{2\pi j a}{\xi_m}\right)\right] , 
    \tanh\left[\lambda \cos\left(\frac{2\pi j a}{\xi_m}\right)\right] \right),
\end{equation}
with the parameter $\lambda$ controlling the length of the magnetic domains $d$ and domain walls $w$.

We further discretize the model in \cref{eq:ty-domain,eq:ty-domain-app} to gain more control over its parameters. To do so, we divide all the indices along the junction into \textit{domains} and \textit{domain walls}. The domains are regions where the magnetization is constant, and the domain walls as smaller regions ($w\leq d$ in what follows) where the magnetization rotates speedily. The domain walls thus represent a region where the exchange interaction between domains with opposite magnetization forces the spins to rotate between those orientations. 

In a junction of width $N_ya$ there are $n=\lfloor N_ya/\xi_m \rfloor$ full periods, with $\lfloor x \rfloor$ being the integer floor of the real number $x$. There are thus at least $2n$ domains and domain walls on each width. For simplicity, we assume that the first interface site $(i,j)=(N_x/2,1)$ always starts a domain so that sites belonging to the $k$-th domain are in the interval 
\begin{equation}
    D_k = \left[ 1+ (k-1)(d+w) , d + (k-1)(d+w) \right] ,
\end{equation}
and those in the $k$-th domain wall belong to
\begin{equation}
    W_k = \left[ 1+ d+ (k-1)(d+w) , k(d+w) \right] .
\end{equation}

The magnetization of the domain model is thus given by 
\begin{equation}
    \bm{t}_j =
    \begin{cases}
        (-1)^{k-1}t_m (0, 0, 1), & j\in D_k 
        \\ 
        t_m \left[ 
        0, 
        \sin\left( x_k + j_w \right), 
        \cos\left( x_k + j_w \right)
        \right], & j\in W_k 
    \end{cases},
    \label{eq:domain_model}
\end{equation}
with 
\begin{equation}
    x_k = \frac{1+(-1)^k}{2} \pi , \quad \text{and} \quad j_w=  \frac{j -kd + (k-1)w}{w+1} \pi .
\end{equation}

We use the discretized model in \cref{eq:domain_model} to compute the supercurrent in \cref{fig3}(g-i) of the main text and the phase diagram in \cref{fig:analytic_numeric_comp} below. 


\subsection{Generalized domain model for disorder effects. }

To introduce disorder effects we can further generalize the domain model in \cref{eq:domain_model} by allowing each domain and domain wall to have a different size. Sites belonging to the $k$-th domain of length $d_k$ are then in the interval
\begin{equation}
    D_k = [1+\sum_{l=1}^{k-1}(d_l+w_l),d_k + \sum_{l=1}^{k-1}(d_l+w_l)],
\end{equation}
and those in the $k$-th domain wall of length $w_k$ are inside
\begin{equation}
    W_k = [1+d_k+\sum_{l=1}^{k-1}(d_l+w_l), \sum_{l=1}^{k}(d_l+w_l)].
\end{equation}

In the following, we do not randomize the length of domain walls, $w_k$, assuming that in physical setups the exchange interaction that rotates the spins between neighboring domains is fixed as long as the domains are sufficiently large. 

The magnitude of the magnetization in each domain can also be different and we, therefore, write
\begin{equation}
    \bm{t}_j =
    \begin{cases}
        t_m^{k} (0, 0, 1), & j\in D_k\\
        \mbf{t}_{m,W}^{k}(j), & j\in W_k\\
    \end{cases},
    \label{eq:domain_model_general}
\end{equation}
where $t_m^k$ is the magnetic tunnelling through the $k$-th domain. The magnitude of the $k$-th domain wall tunneling, $\mbf{t}_{m,W}^k$, must now vary spatially to adjust to the different magnitudes of the neighboring domains $k$ and $k+1$. We thus have
\begin{equation}
    \mbf{t}_{m,W}^k(j) = \left(t_m^k\frac{j_e^k-j+1/2}{j_e^k-j_s+1} + t_m^{k+1}\frac{j-j_s^k+1/2}{j_e^k-j_s^k+1}\right) \left[0, s_r^k\sin(x_k), \cos(x_k) \right].
\end{equation}
Here, $j_s^k = 1+d_k+\sum_{l=1}^{k-1}(d_l+w_l)$ is the index of the starting site in the $k$-th domain wall and $j_e^k=\sum_{l=1}^{k}(d_l+w_l)$ is that of the ending site. We introduced the parameter $s_r^k$ to control the ``sign of the rotation''. For example, in the harmonic model it is set to $1$. 

According to these expressions:
\begin{itemize}
    \item A perfectly antiferromagnetic order of domains is imposed by $\sgn(t_m^k) = -\sgn(t_m^{k-1})$.
    \item A ``domain flip'' disorder is introduced using $\sgn({t_m^{k}}) = n_{\rm df}^k\sgn({t_m^{k-1}})-(1-n_{\rm df}^k)\sgn({t_m^{k-1}})$, where at the $k$-th domain there is a chance $\sigma_{\rm df}$ that $n_{\rm df}^k = 1$ (otherwise 0), in which case the domains $k$ and $k+1$ are ferromagnetically aligned.
    \item ``Domain wall flip'' disorder is set by choosing $s_r^k$ from a uniform random distribution in the interval $(-\sigma_{\rm dwf}, 1-\sigma_{\rm dwf})$. Therefore, at the $k$-th domain wall the ``sign of the rotation'' is allowed to change with probability $\sigma_{\rm dwf}$. 
    \item Disorder in the magnetic amplitude, the domain size, or the electrostatic potential are introduced by randomly choosing the values of $|t_m^{k}|$, $d_k$, or $\mu_{i,j}$ from a Gaussian distribution of mean $\bar{t}_m$, $\bar{d}$, or $\bar{\mu}$ and standard deviation $\sigma_{t_m}$, $\sigma_{\xi_m}$, or $\sigma_{\mu}$, respectively. 
\end{itemize}


\begin{figure}[t]
 \centering
 \subfloat{\includegraphics[width=.65\textwidth]{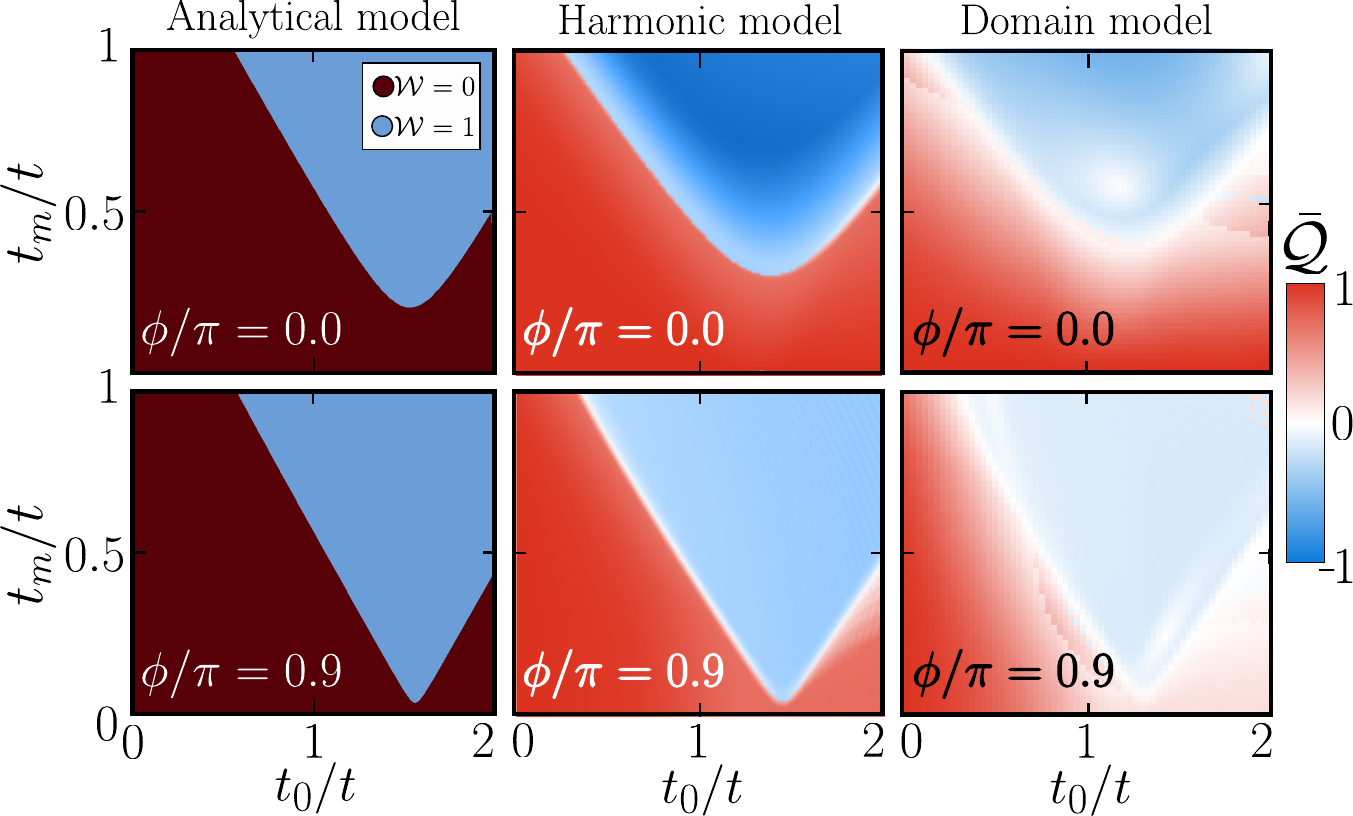}}
 \caption{Topological phase diagram as a function of $t_0$ and $t_{m}$ for $\phi=0$ (top panels) and $\phi=0.9\pi$ (bottom panels). 
 The left column shows the winding number $\mathcal{W}$ for the 1D model with $\mu_\text{eff}/t=-1.85$. 
 In the central column we compute the parity $\bar{\mathcal{Q}}$ of the approximate topological invariant for a finite size system with a harmonic texture and parameters $N_x=16$, $N_y=256$, and $\mu/t = -3.85$. In the right column we show the domain texture with $N_x=32$, $N_y=512$, $w=1$, $d=2$, and $\mu/t=-3.35$. }
 \label{fig:analytic_numeric_comp}
\end{figure}

\section{Approximate topological invariant\label{sec:app-aprox-top-inv}} 

The topological invariant, \cref{eq:topo-inv}, is only well defined for infinite systems. By analogy with the analytical model, \cref{fig:analytic_numeric_comp}, we now define an approximate topological invariant to help us distinguish the presence of zero-energy edge modes in the finite-size system. 

We first compute the eigenvalues $\epsilon_n$ satisfying $H \ket{\psi^{\pm}} = \pm \epsilon_n \ket{\psi^{\pm}}$ for the eigenvectors $\ket{\psi^{\pm}}$ of the Hamiltonian $H=H_L+H_t+H_R$ in the main text (see \cref{eq:hlr,eq:vlr} in the main text). 
In our description, $\bar{\mathcal{W}}$ quantifies the number of localized \glspl{mbs} pairs, $ \{ \pm\epsilon_1, \pm\epsilon_2, ..., \pm\epsilon_{\bar{\mathcal{W}}} \} $, lying at $E \approx 0$ well separated from the bulk of states. Consequently, $\bar{\mathcal{W}}$ resembles the winding number characterizing the topological phase for an infinite size system. 

The parity of this pseudo-winding number, $\bar{\mathcal{M}}=(-1)^{\bar{\mathcal{W}}}$, determines the even or odd number of pairs of \glspl{mbs} at the edges of the system, while the size of the topological gap is simply $\epsilon_{\bar{\mathcal{W}}+1}$.
In the simplest scenario, with a single eigenstate close to zero energy, we have
\begin{equation}\label{eq:approx-top-inv}
\bar{\mathcal{M}}=(-1)^{\bar{\mathcal{W}}} = 
     \begin{cases}
        1, &\epsilon_2 / 2 < \epsilon_1 \quad\text{(trivial)}\\
        -1, &\epsilon_2 / 2 > \epsilon_1 \quad\text{(nontrivial)}
     \end{cases} ,
\end{equation}
while the gap is given by
\begin{equation}\label{eq:top-gap}
\delta = 
     \begin{cases}
        \epsilon_1, &\epsilon_2 / 2 < \epsilon_1 \quad\text{(trivial gap)}\\
        \epsilon_2, &\epsilon_2 / 2 > \epsilon_1 \quad\text{(nontrivial gap)}
     \end{cases} .
\end{equation}

We can now study the topology of a junction with finite-size superconductors and compare it to the analytical one-dimensional model in \cref{fig:analytic_numeric_comp}. In the left column we show $\mathcal{M}=(-1)^{\mathcal{W}}$ according to \cref{eq:M} for the analytical model. For the finite-size system we compute the parity of the approximate topological invariant $\bar{\mathcal{M}}$ in \cref{eq:approx-top-inv} re-scaled by the topological gap in \cref{eq:top-gap}, i.e., $\bar{\mathcal{Q}}=\bar{\mathcal{M}} \left(\delta/\Delta_0\right)$. 
The blue regions in the central and right columns of \cref{fig:analytic_numeric_comp} thus represent the topologically nontrivial phases, with the boundary where the gap closes appearing in white. 
It is clear that the phase diagram for the analytical model is qualitatively the same as for the the finite-size junction with a harmonic texture (central column of \cref{fig:analytic_numeric_comp}) or a domain one (right column on \cref{fig:analytic_numeric_comp}). The latter features extra low-energy trivial states that can lead to non-topological gap closings for some parameters. That is the case for the white lines in the right column of \cref{fig:analytic_numeric_comp} that separate two trivial red regions. 

\begin{figure}[t]
 \centering
 \subfloat{\includegraphics[width=.65\linewidth]{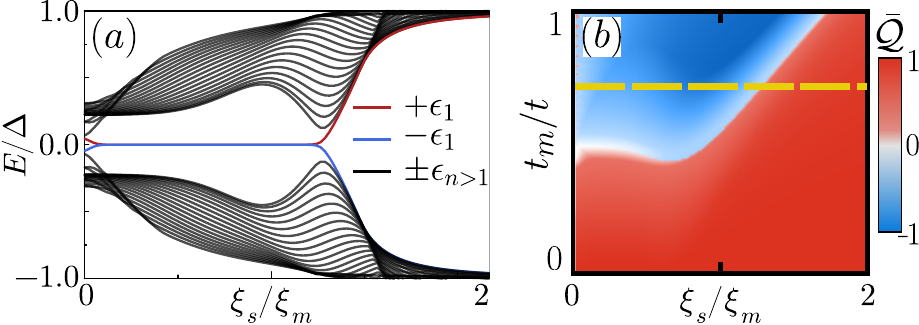}}
  \caption{(a) Lowest energy bands as a function of $\xi_{m}$ for the yellow line in panel (b). The map in panel (b) is repeated from \cref{fig2}(d) of the main text. }
 \label{fig:bands-xi_m}
\end{figure}

We corroborate the predictions of the approximate topological invariant studying the energy spectrum. In our calculations for finite-size junctions we can vary $\xi_m$ from $2a$ (purely antiferromagnetic order at the interface) to $\xi_m\to\infty$ (pure ferromagnetic order). 
We show a representative result in \cref{fig:bands-xi_m}(a) illustrating the onset of lowest-energy edge states and the emergence of zero energy states for $\xi_m\sim\xi_s$. 
This calculation corresponds to a fixed value of $t_m$ in \cref{fig2}(d) of the main text, see yellow line of \cref{fig:bands-xi_m}(b). 

By contrast, the system is in the trivial regime when these two length scales are very different. For $\xi_m\gg\xi_s$ the superconductor feels a constant exchange field locally that can close the superconducting gap for sufficiently large $t_m$ values. On the other hand, for $\xi_m\ll\xi_s$ the superconductor averages out the barrier magnetization recovering the conventional superconducting spectrum. 



\begin{figure}[b]
 \centering
 \subfloat{\includegraphics[width=\linewidth]{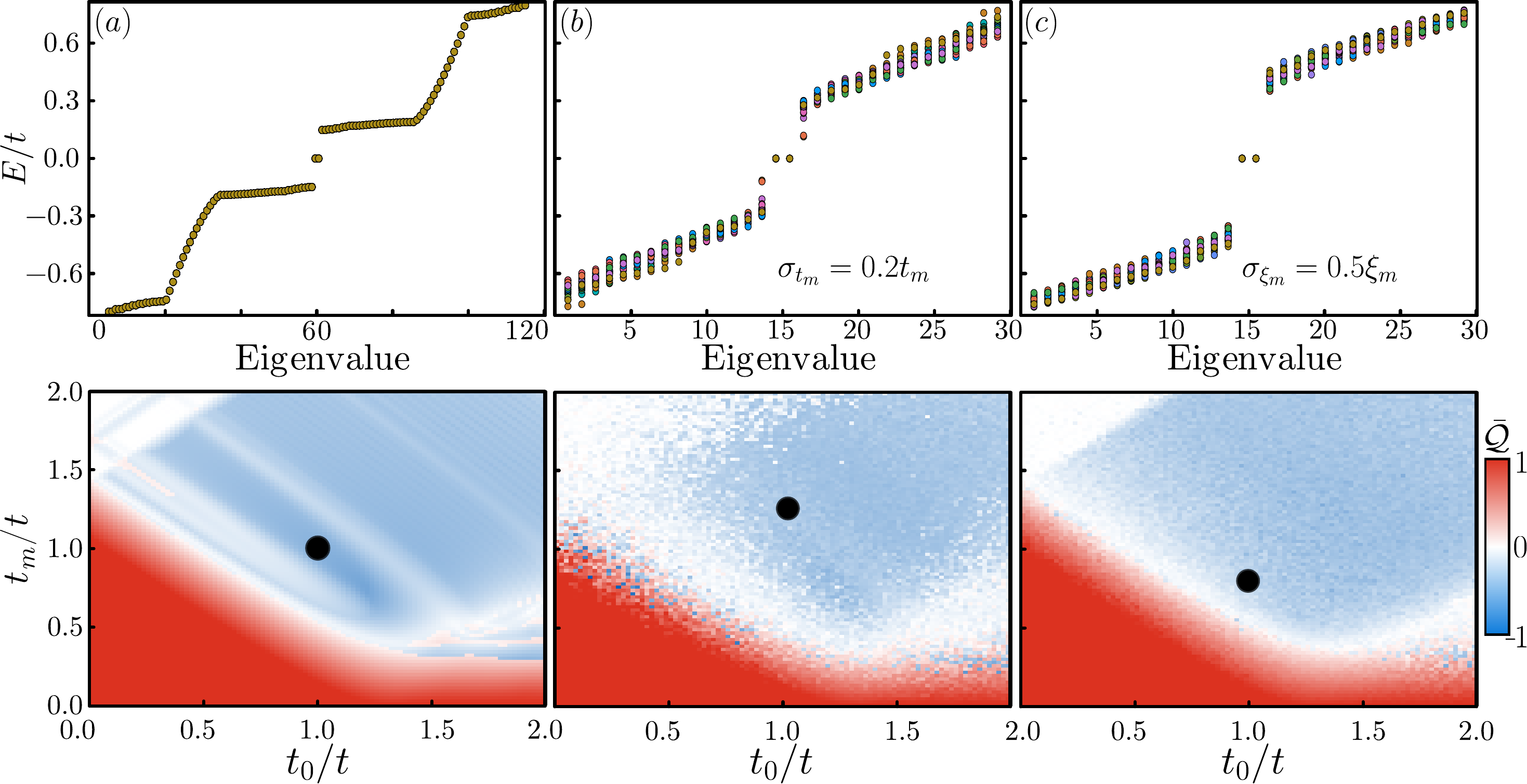}}
 \caption{Topological phase diagram for the domain model showing the energy distribution of eigenvalues (top panels) and the approximate invariant $\bar{\mathcal{Q}}$ (bottom panels). We show the case without disorder in (a) to compare it with the effect of disorder in the magnitude $t_m$ (b) and the period $\xi_m$ (c) of the magnetic texture. 
 Parameters are $d=6$, $w=2$, $\mu/t=4$, $\Delta_0/t=0.2$, $t_0/t=1$, $N_x=8$, and $N_y=100$.
 }
 \label{fig:noise-maps_a}
\end{figure}

\section{Role of disorder\label{sec:app-disorder}}

The minimal model in \cref{sec:app-analytics} has helped us connect the magnetic spatial variation along the interface with the emergence of a synthetic spin-orbit coupling responsible for the nontrivial topology. We have then shown that the topological phase transition is robust when this idealized model is extended to a finite-size sequence of domain regions, see \cref{fig:analytic_numeric_comp}. We now study the effect of disorder in the domain model to test the stability of the topological phase, see \cref{sec:app-domain-model}. 

Starting with a perfectly ordered sequence of domains connecting finite-size superconductors, see \cref{fig:noise-maps_a}(a), we first consider the possibility of having spatial irregularities in the period and the magnitude of the magnetic texture. 
Disorder effects are thus included by modifying the value of a specific parameter $p$ using a Gaussian distribution with a standard deviation $\sigma_p$, e.g., $p=t_m$ in \cref{fig:noise-maps_a}(b). 
Each map point in the bottom panels of \cref{fig:noise-maps_a} then represents a random realization with the indicated $\sigma_p$. On the top panels we show the lowest energy eigenvalues for $20$ random realizations using the parameters indicated by black dots on the corresponding maps. 

\begin{figure}[t]
 \centering
 \subfloat{\includegraphics[width=.75\linewidth]{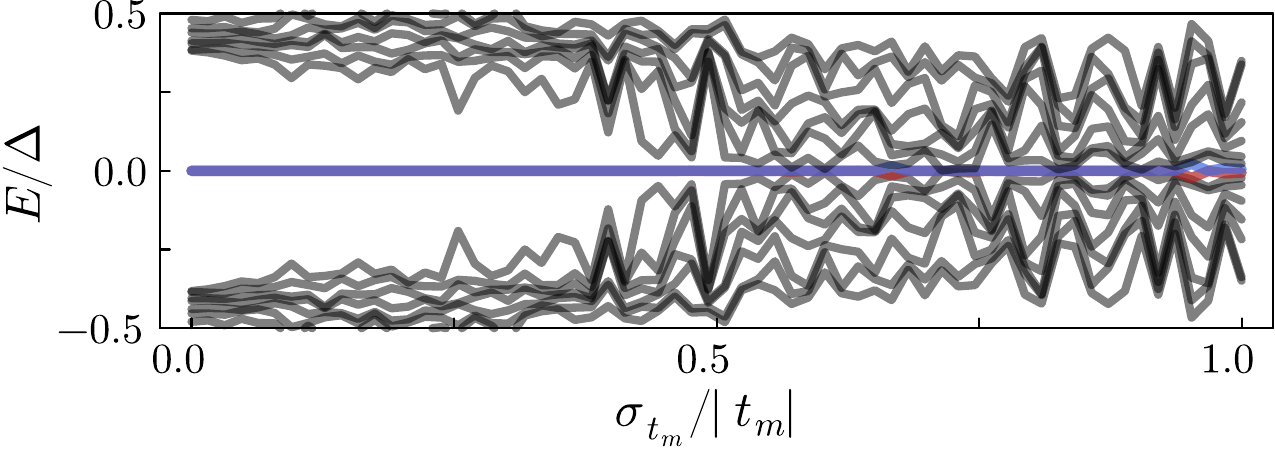}}
 \caption{Lowest energy bands (gray lines) and zero energy mode (blue line) in the topological phase as a function of the disorder strength in the magnitude of the magnetization $t_m$. Parameters are the same as in \cref{fig:noise-maps_a}(b) with $\bar{t}_m/t=1.2$. }
 \label{fig:tm-noise}
\end{figure}

The boundary of the topological phase is sensitive to the amplitude of the magnetization, $t_{m}$, as discussed above and shown in \cref{fig:analytic_numeric_comp} and \cref{fig2} of the main text. 
Therefore, a disorder on the magnitude of $t_m$ blurs the critical boundary of the topological phase and also reduces the topological gap inside it, see the white region in the bottom panel of \cref{fig:noise-maps_a}(b). The topological gap reduction can be seen in the eigenvalue distribution on the top panel; for this case, the topological gap is still open. 
In order to fully close the gap we need to introduce a disorder strength comparable to the minimal magnetic amplitude required to enter the topological phase. For example, \cref{fig:tm-noise} indicates that the topological gap for $t_m/t\sim1$ is closed when the variations in amplitude reach $50\%$ of the amplitude without disorder. Note, however, that the system is still in the nontrivial phase with zero energy modes indicated by the colored lines in \cref{fig:tm-noise}. 

One of our main results is that the topological phase is easier to achieve when the magnetic and the superconducting coherence lengths are comparable, i.e., $\xi_m\sim\xi_s$. 
By introducing disorder on the domains forming each period $\xi_m$, while keeping the domain wall size fixed, we can analyze the effect of having domains with different sizes. The phase diagram in \cref{fig:noise-maps_a}(c) indicates that the topological region is barely affected by this type of disorder. 

\begin{figure}[b]
 \centering
 \subfloat{\includegraphics[width=\linewidth]{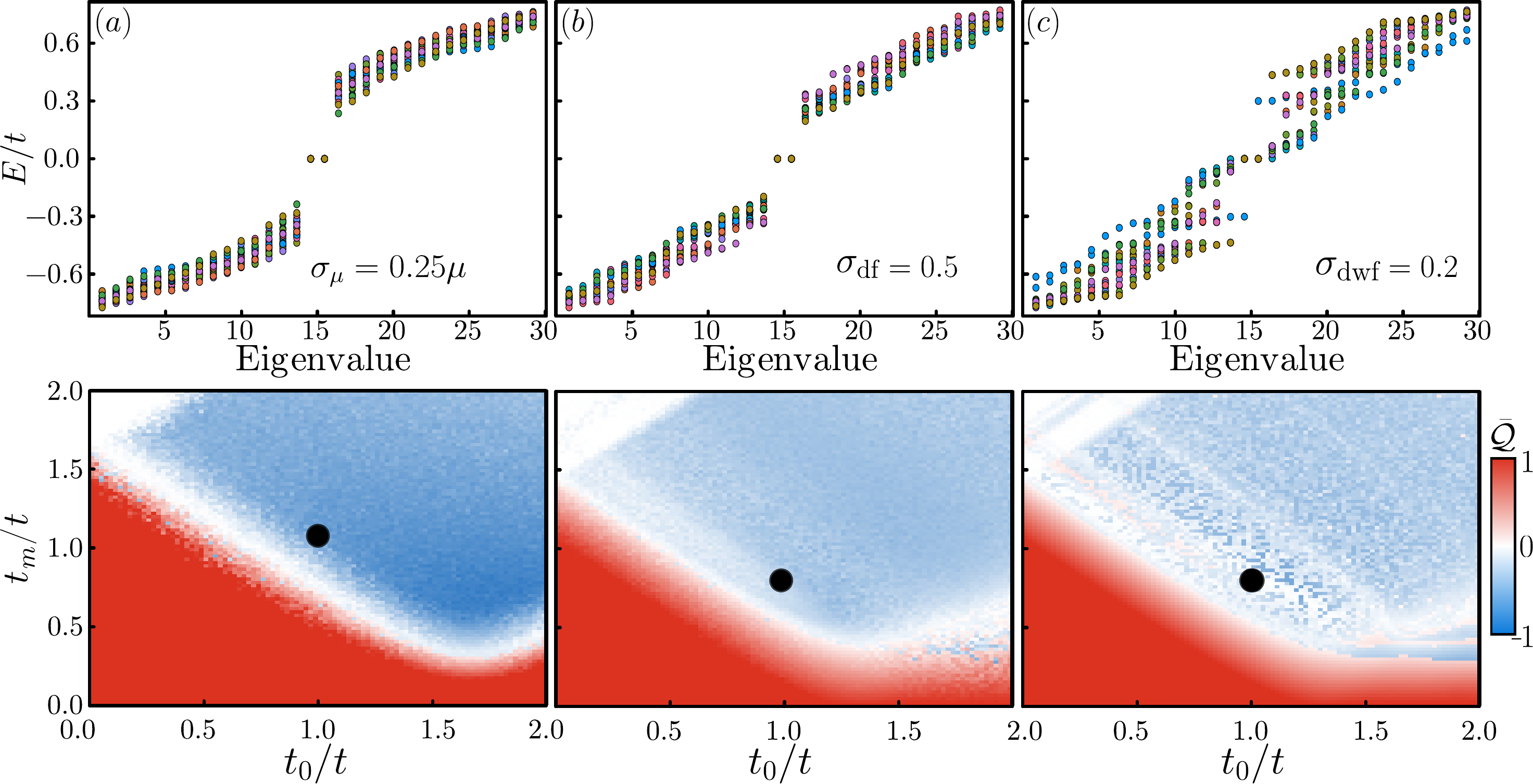}}
 \caption{Topological phase diagram for the domain model and lowest energy eigenvalues showcasing disorder effects on (a) the electrostatic potential, (b) the magnetic orientation of the domains, and (c) the orientation of the domain walls. Parameters are $d=6$, $w=2$, $\mu/t=4$, $\Delta_0/t=0.2$, $t_0/t=1$, $N_x=8$, and $N_y=100$. }
 \label{fig:noise-maps_b}
\end{figure}

We can also study non-magnetic disorder effects introducing a random variation on the chemical potential $\mu$. We consider in \cref{fig:noise-maps_b}(a) a disorder strength in $\mu$ that can, in principle, reach values that would take the system out of the topological region, see \cref{fig2}(a) in the main text. However, the phase diagram is almost unaffected and the topological gap barely reduced. The topological phase is thus robust to small changes in the band filling. 

Finally, we consider effects that would directly affect the emerging synthetic spin-orbit coupling at the interface. 
First, in \cref{fig:noise-maps_b}(b) we introduce a random flip in the orientation of each domain, that is, in the sign of $t_m$ for each group of parallel spins in a domain, see \cref{eq:domain_model,eq:domain_model_general}. This disorder, however, does not affect the sign of $t_m$ in the domain walls. As a result, the topological gap is reduced a bit, but the nontrivial phase is maintained. 

By contrast, the opposite situation where the disorder randomly changes the rotation in the domain walls has a stronger effect on the topological phase. The map in \cref{fig:noise-maps_b}(c) is computed imposing domain-wall-flips around 80\% of the times and features many white regions inside the nontrivial area where the topological gap has closed. This type of disorder is affecting the rotation direction of the magnetic texture and, therefore, the effective spin-orbit coupling. 
When the direction of the magnetization rotation changes the average spin-orbit is zero locally at one point, setting a topological phase boundary and leading therefore to additional Andreev bound states inside the gap. If this happens randomly along the sample with enough frequency many trivial low-energy states will localize at the junction, leading to a dense density of states around zero energy that could be a false positive in experimental setups~\cite{Burset_2021}. 

In summary, variations in the magnetization magnitude or period and fluctuations on the exchange field are not enough to drive the system from the topological to the trivial phase; although they can reduce the topological gap. By contrast, it is important to maintain the  magnetization rotation to avoid the appearance of undesired low-energy states inside the system.

\begin{figure}[t]
 \centering
 \subfloat{\includegraphics[width=1\textwidth]{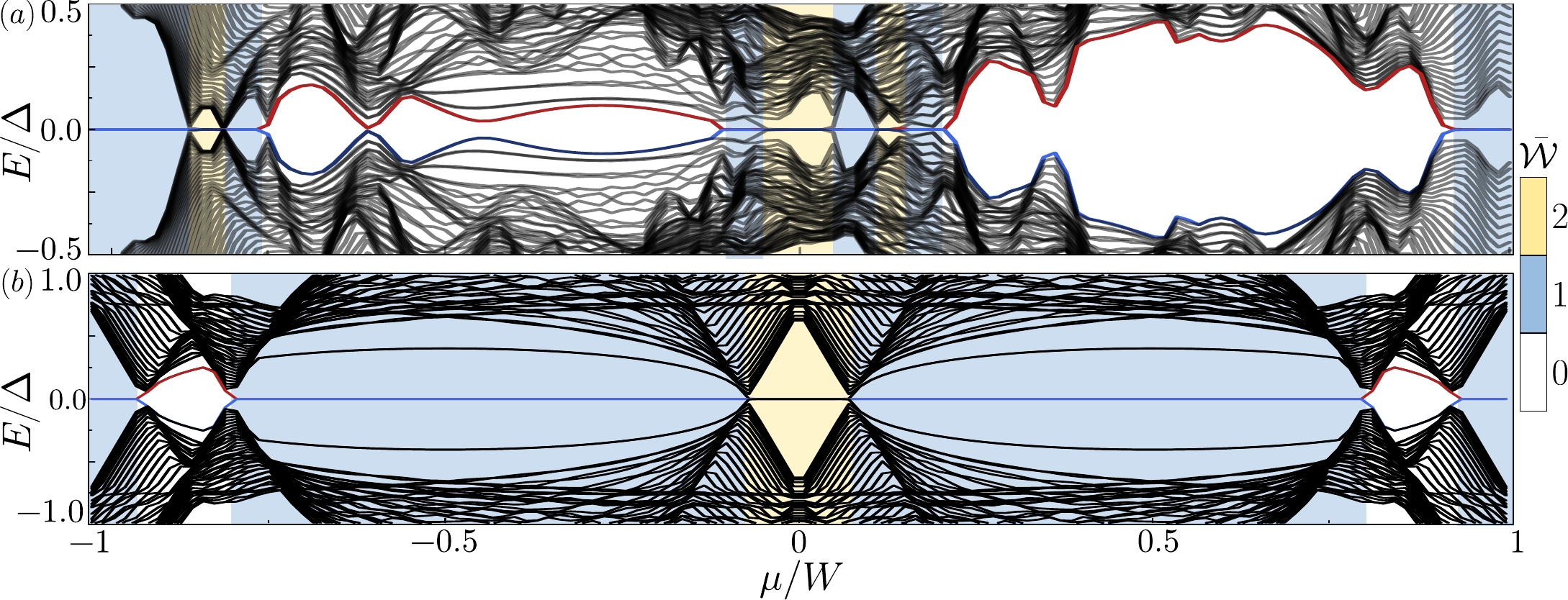}}\\
 \caption{Energy bands as a function of $\mu$ (in units of the bandwidth $W$) for (a) the finite size system with the same parameters as in \cref{fig2} of the main text and (b) the analytic 1D model, corresponding to $N_x=2$. }
 \label{fig:all_fillings}
\end{figure}

\section{Band filling\label{sec:app-band-fill}}

In the main text we focused on chemical potentials in the region $\mu\sim-3.5t$, which maintain electron-hole symmetry in the two dimensional bands and feature only one pair of \gls{mbs} in the topological phase. However, as we show in \cref{fig2}(a) of the main text, different topological phases emerge as a function of $\mu$. We now show in \cref{fig:all_fillings}(a) the full range of phases for a setup with the same parameters as \cref{fig2} of the main text. The lowest energy states are showcased as red and blue lines for $\epsilon_1$ and $-\epsilon_1$, respectively. 

Similarly, the analytical 1D model can be approximated to a finite length chain setting $N_x = 2$ and a finite $N_y$. With two edges in the $y$ direction, we can explore the emergence of topological zero-energy states in the 1D model. \Cref{fig:all_fillings}(b) shows that the topological phase is more robust than in the bulk case, extending for almost every value of $\mu$ for a sufficiently large $t_m/t$ value. For chemical potentials around $\mu\sim0$ we find topological phases characterized by more than one Majorana state at each edge in both models (see yellow regions in \cref{fig:all_fillings}).

\section{Extended magnetic texture\label{sec:app-ext-mag-texture}}

To represent a Josephson junction mediated by a finite-size magnetic texture we couple the superconducting leads, \cref{eq:hlr} in the main text, to a central region described by
\begin{align}\label{eq:finite-Hamil}
 H_{C} ={}& -\sum_{\sigma=\up,\dw} 
 ( t_0\sum_{ \langle x, x' \rangle } c_{i'j',\sigma}^\dagger c_{ij,\sigma} + \sum_{i,j} \mu_{ij} c_{ij,\sigma}^\dagger c_{ij,\sigma} 
 )
  + \sum_{\substack{i, j\\ \alpha\beta}} c^\dagger_{ij,\alpha} \left( \bm{m}_j \cdot \bm{\sigma} \right)_{\alpha,\beta} c^\dagger_{ij,\beta} ,
\end{align}
where the index $i$ runs from $1$ to $N_W$, the number of barrier sites in the transport direction. 
Note that we now model the magnetic texture as a local magnetization with constant magnitude, $|\mathbf{m}_j|=m$, as
\begin{equation}\label{eq:my-sin}
    \bm{m}_j = m \left[\sin{(2\pi j a/\xi_m)}, 0, \cos{(2\pi j a/\xi_m)}\right].
\end{equation}
The Hamiltonian of the coupled system is
\begin{equation}
H =
     \begin{pmatrix}
      H_L & V_{LC} & 0 \\
      V_{LC}^\dagger & H_C & V_{RC}^\dagger \\
      0 & V_{RC} & H_R
     \end{pmatrix} ,
\end{equation}
where
\begin{equation} \label{eq:app-vlr}
    V_{XC} = - \frac{t_0}{2} \sum_{j,\sigma} \hat{c}^\dagger_{x,j,\sigma} \hat{c}_{x+1,j,\sigma} + \mathrm{h.c.}\,, 
\end{equation}
with $x=N_x/2$, the width of the left superconductor region, when $X=L$ in $V_{LC}$, and $x=N_x/2+N_W+1$ when $X=R$ in $V_{RC}$. 

\begin{figure}[t]
 \centering
\subfloat{\includegraphics[width=.55\linewidth]{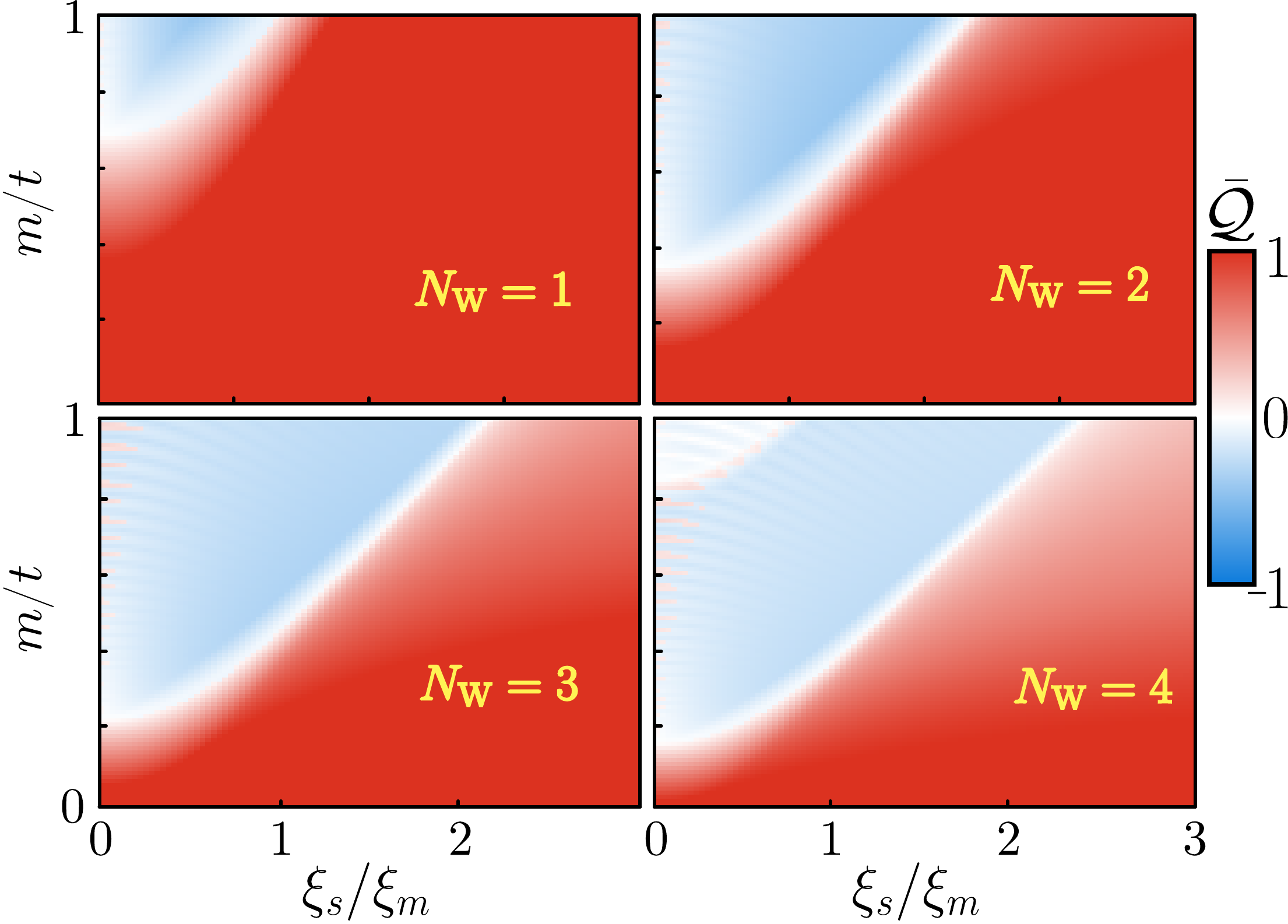}}
 \caption{Topological phase diagram of a long Josephson junction including $N_W$ central magnetic-textured sites between superconductors. The rest of parameters are $\Delta_0/t = 0.2$, $t_0 / t= 1$ and $\mu = -3.85t$. }
 \label{fig:interfaces}
\end{figure}

In \cref{fig:interfaces} we show the topological phase diagram as a function of $m$ and $\xi_m$ for different widths of the central region, $N_W=1,\dots,4$. As the barrier width is increased, the topological gap is reduced due to the lower transmission. However, the minimum $t_m $ to achieve the topological phase is also minimised, approaching the same value $\Delta_0$ as the one predicted for the bulk minimal model. 

\section{Current-phase relation\label{sec:app-cpr}}
Finally, we describe here the calculation of the Josephson current for a phase-biased junction shown in \cref{fig3} of the main text. 

From diagonalization of the full Hamiltonian, \cref{eq:hlr,eq:vlr} in the main text, we obtain a set of eigenvalues $E_i(\phi)$. The corresponding free energy at a temperature $T$ can be computed as
\begin{equation}
	F = -T \log(Z) = -8 k_BT \sum_i \log\left[\cosh\left(\dfrac{E_i}{2k_BT}\right)\right] ,
\end{equation}
where $Z=\sum_i\e^{-E_i/{k_BT}}$ is the partition function of the system and $k_B$ the Boltzmann constant. 

We define the Josephson current \cite{Zagoskin_Springer1998} as the derivative with respect to the superconducting phase difference $\phi$, namely, 
\begin{equation}
	I(\phi) = 2e \dfrac{dF}{d\phi} = -8e\sum_i \tanh \left(\dfrac{E_i}{2k_BT} \right) \dfrac{dE_i}{d\phi} .
\end{equation}
In the limit of zero temperature the supercurrent simply reduces to
\begin{equation}
    I(\phi) = -8e\sum_{E_i<0}\frac{\partial E_i}{\partial \phi}.
\end{equation}

\end{document}